\documentclass[preprint,amsmath,amssymb,aps,showkeys,showpacs]{revtex4}
\usepackage[english]{babel}
\usepackage{graphicx}
\usepackage{graphics}
\usepackage{amsmath}
\usepackage{dcolumn}
\usepackage{amssymb}
\usepackage{bm}
\usepackage[latin1]{inputenc}

\begin{document}
\title{On a relativistic scalar particle subject to the Klein-Gordon oscillator, the Coulomb potential and a linear scalar potential}
\author{R. L. L. Vit\'oria}
\affiliation{Departamento de F\'isica, Universidade Federal da Para\'iba, Caixa Postal 5008, 58051-970, Jo\~ao Pessoa, PB, Brazil.} 
\author{C. Furtado}
\email{furtado@fisica.ufpb.br} 
\affiliation{Departamento de F\'isica, Universidade Federal da Para\'iba, Caixa Postal 5008, 58051-970, Jo\~ao Pessoa, PB, Brazil.} 
\author{K. Bakke}
\email{kbakke@fisica.ufpb.br} 
\affiliation{Departamento de F\'isica, Universidade Federal da Para\'iba, Caixa Postal 5008, 58051-970, Jo\~ao Pessoa, PB, Brazil.}

\begin{abstract}
The relativistic quantum dynamics of an electrically charged particle subject to the Klein-Gordon oscillator and the Coulomb potential is investigated. By searching for relativistic bound states, a particular quantum effect can be observed: a dependence of the angular frequency of the Klein-Gordon oscillator on the quantum numbers of the system. The meaning of this behaviour of the angular frequency is that only some specific values of the angular frequency of the Klein-Gordon oscillator are permitted in order to obtain bound state solutions. As an example, we obtain both the angular frequency and the energy level associated with the ground state of the relativistic system. Further, we analyse the behaviour of an electrically charged particle subject to the Klein-Gordon oscillator, the Coulomb potential and a linear scalar potential.  

\end{abstract}


\maketitle

\section{Introduction}

Relativistic effects on quantum systems where the motion of a particle is governed by harmonic oscillations have been the subject of several works in the literature \cite{bah,bah2,bah3,bah4}. Examples of these studies are the binding of heavy quarks \cite{qui,chai}, the vibrational spectrum of diatomic molecules \cite{ikh} and oscillations of atoms in crystal lattices by mapping them as a position-dependent mass system \cite{pdm,pdm2,pdm3,pdm4}. In recent decades, the relativistic generalization of the harmonic oscillator has attracted a great deal of attention whose model proposed by Moshinsky and Szczepaniak \cite{osc1} is the best known for Dirac particles. This model is called the Dirac oscillator and has a great interest in studies of Jaynes-Cummings model \cite{jay2,osc3}, quantum phase transitions \cite{extra2,extra3} and the Ramsey-interferometry effect \cite{osc6}.

Another interesting model was proposed by Bruce and Minning \cite{kgo} where an analogous coupling to the Dirac oscillator coupling \cite{osc1} is introduced into the Klein-Gordon equation in such a way that one can recover the Schr\"odinger equation for a harmonic oscillator in the nonrelativistic limit. This coupling is called the Klein-Gordon oscillator \cite{kgo,kgo2,kgo8,kgo7,kgo9}. In recent years, the Klein-Gordon oscillator has been investigated in noncommutative space \cite{kgo3,kgo4}, in noncommutative phase space \cite{kgo5} and in $\mathcal{PT}$-symmetric Hamiltonian \cite{kgo6}. In particular, the isotropic Klein-Gordon oscillator in $\left(2+1\right)$ dimensions allows us the write the Klein-Gordon equation in the form:
\begin{eqnarray}
\left[\mathcal{E}^{2}-m^{2}\right]\phi=\left[\hat{p}+im\omega\rho\,\hat{\rho}\right]\cdot\left[\hat{p}-im\omega\rho\,\hat{\rho}\right]\phi,
\label{kgo}
\end{eqnarray}
where $m$ is the rest mass of the scalar particle, $\omega$ is the angular frequency of the Klein-Gordon oscillator, $\rho=\sqrt{x^{2}+y^{2}}$ is the radial coordinate and $\hat{\rho}$ is a unit vector in the radial direction.

In recent years, the relativistic quantum dynamics of a scalar particle subject to different confining potentials has been investigated in several areas of physics \cite{alvaro,qian,castro,alhardi,adame,xu}. A special case is the confinement of a relativistic scalar particle to the Coulomb potential \cite{kg,kg2,kg3,kg4,greiner}. The procedure in introducing a scalar potential into the Klein-Gordon equation occurs by modifying the momentum operator $\hat{p}_{\mu}=i\partial_{\mu}$ in the form: $\hat{p}_{\mu}\rightarrow \hat{p}_{\mu}-q\,A_{\mu}\left(x\right)$ \cite{greiner}. On the other hand, one can also introduce a scalar potential (non-electromagnetic potential) in the Klein-Gordon equation by making a modification in the mass term in the form: $m\rightarrow m+S\left(\vec{r},\,t\right)$, where $S\left(\vec{r},\,t\right)$ is the scalar potential \cite{scalar,greiner}. This modification in the mass term has been explored in recent decades, for instance, by analysing the behaviour of a Dirac particle in the presence of a static scalar potential and a Coulomb potential \cite{scalar2}, and a relativistic scalar particle in the cosmic string spacetime \cite{eug}. Another example is the quark-antiquark interaction discussed in Ref. \cite{bah}, where this interaction is mapped into a problem of a relativistic spinless possessing a position-dependent mass, where the mass term acquires a contribution given by a interaction potential that consists of a linear and a harmonic confining potential plus a Coulomb potential term.

In this work, we investigate the relativistic quantum dynamics of an electrically charged particle under the influence of the Klein-Gordon oscillator and the Coulomb potential. The Coulomb potential is introduced into the Klein-Gordon equation via minimal coupling, that is, as the component $A_{0}$ of the gauge potential. Then, we show that relativistic bound state solutions to the Klein-Gordon equation can be obtained for both attractive and repulsive Coulomb potentials. Moreover, a quantum effect characterized by a restriction of the values of the angular frequency of the Klein-Gordon oscillator is investigated. We show that this restriction is imposed by the quantum numbers of the system. From the mathematical point of view, this dependence of the angular frequency of this relativistic oscillator on the quantum numbers results from obtaining a polynomial solution to the biconfluent Heun equation. From the quantum mechanics point of view, this is an effect which arises from the influence of the Coulomb potential on the Klein-Gordon oscillator, whose meaning is that not all values of the angular frequency are allowed. Besides, we analyse the relativistic quantum dynamics of an electrically charged particle under the influence of the Klein-Gordon oscillator, the Coulomb potential and a linear scalar potential.

The structure of this paper is as follows: in section II, we study the relativistic quantum dynamics of a scalar particle under the influence of the Klein-Gordon oscillator and the Coulomb potential in the Minkowski spacetime in $\left(2+1\right)$ dimensions. We obtain relativistic bound state solutions to the Klein-Gordon equation for both attractive and repulsive Coulomb potentials. Besides, we analyse a quantum effect characterized by a dependence of the angular frequency of this relativistic oscillator on the quantum numbers; in section III,  we analyse the relativistic quantum dynamics of an electrically charged particle under the influence of the Klein-Gordon oscillator, the Coulomb potential and a linear scalar potential; in section IV, we present our conclusions.

\section{Klein-Gordon oscillator under the influence of a Coulomb potential}

In this section, we investigate the relativistic quantum dynamics of a scalar particle under the influence of the Klein-Gordon oscillator and the Coulomb potential in (2+1) dimensions. We analyse the influence of the Coulomb potential on the spectrum of energy of the Klein-Gordon oscillator. By following Ref. \cite{greiner}, the Coulomb potential is introduced into the Klein-Gordon equation via minimal coupling, that is, as the component $A_{0}$ of the gauge potential. Therefore, the Klein-Gordon equation with an electromagnetic field is written by introducing the minimal coupling $\hat{p}_{\mu}-q\,A_{\mu}$, where $q$ is the electric charge and $A_{\mu}=\left(-A_{0},\,\vec{A}\right)$ is the electromagnetic 4-vector potential. Therefore, we have \cite{greiner}:
\begin{eqnarray}
\left[\hat{p}^{\mu}-q\,A^{\mu}\right]\left[\hat{p}_{\mu}-q\,A_{\mu}\right]\phi-m^{2}\,\phi=0,
\label{1.1}
\end{eqnarray}
where the line element of the Minkowski spacetime is given in the form:
\begin{eqnarray}
ds^{2}=-dt^{2}+d\rho^{2}+\rho^{2}\,d\varphi^{2},
\label{1.2}
\end{eqnarray}
and the Coulomb potential is given by 
\begin{eqnarray}
q\,A_{0}=\frac{f}{\rho}=\pm\frac{\left|f\right|}{\rho},
\label{1.3}
\end{eqnarray}
where $f$ is a constant and  $\rho=\sqrt{x^{2}+y^{2}}$ is the radial coordinate.

Henceforth, let us consider a relativistic scalar particle subject to the Coulomb potential (\ref{1.3}) and the Klein-Gordon oscillator given in Eq. (\ref{kgo}). Thereby, the Klein-Gordon equation becomes
\begin{eqnarray}
m^{2}\phi=\left[i\frac{\partial}{\partial t}+q\,A_{0}\right]^{2}\phi-\left[\hat{p}+im\omega\rho\,\hat{\rho}\right]\cdot\left[\hat{p}-im\omega\rho\,\hat{\rho}\right]\phi,
\label{1.4}
\end{eqnarray}
where $m$ is the rest mass of the particle, $\omega$ is the angular frequency of the Klein-Gordon oscillator and $\hat{\rho}$ is a unit vector in the radial direction. Thereby, the Klein-Gordon equation (\ref{1.4}) in the Minkowski spacetime, in (2+1) dimensions, becomes
\begin{eqnarray}
m^{2}\phi=-\frac{\partial^{2}\phi}{\partial t^{2}}+i\frac{2f}{\rho}\frac{\partial\phi}{\partial t}+\frac{f^{2}}{\rho^{2}}\,\phi+\frac{\partial^{2}\phi}{\partial\rho^{2}}+\frac{1}{\rho}\frac{\partial\phi}{\partial\rho}+\frac{1}{\rho^{2}}\frac{\partial^{2}\phi}{\partial\varphi^{2}}+m\omega\,\phi-m^{2}\omega^{2}\rho^{2}\,\phi.
\label{1.5}
\end{eqnarray}

In what follows, let us consider a particular solution to Eq. (\ref{1.5}) which is an eigenfunction of the operator $\hat{L}_{z}=-i\partial_{\varphi}$. Therefore, we can write a particular solution to Eq. (\ref{1.5}) in terms of the eigenvalues of the $z$-component of the angular momentum, $\hat{L}_{z}=-i\partial_{\varphi}$, as follows:
\begin{eqnarray}
\phi\left(t,\,\rho,\,\varphi\right)=e^{-i\mathcal{E}t}\,e^{il\varphi}\,R\left(\rho\right),
\label{5}
\end{eqnarray}
 where $l=0,\pm1,\pm2,\ldots$ and $R\left(\rho\right)$ is a function of the radial coordinate. Then, substituting (\ref{5}) into Eq. (\ref{1.5}), we obtain
\begin{eqnarray}
\frac{d^{2}R}{d\rho^{2}}+\frac{1}{\rho}\frac{dR}{d\rho}-\frac{\gamma^{2}}{\rho^{2}}\,R+\frac{2\mathcal{E}f}{\rho}\,R-m^{2}\omega^{2}\rho^{2}\,R+\beta\,R=0,
\label{6}
\end{eqnarray}
where we have defined the following parameters:
\begin{eqnarray}
\beta&=&\mathcal{E}^{2}-m^{2}+m\omega;\nonumber\\
[-2mm]\label{7}\\[-2mm]
\gamma^{2}&=&l^{2}-f^{2}.\nonumber
\end{eqnarray}

From now on, let us consider $r=\sqrt{m\omega}\,\rho$, thus, we rewrite the radial equation (\ref{6}) in the form:
\begin{eqnarray}
\frac{d^{2}R}{dr^{2}}+\frac{1}{r}\frac{dR}{dr}-\frac{\gamma^{2}}{r^{2}}\,R+\frac{\delta}{r}\,R-r^{2}\,R+\frac{\beta}{m\omega}\,R=0,
\label{8}
\end{eqnarray}
where we have defined a new parameter
\begin{eqnarray}
\delta=\frac{2\mathcal{E}f}{\sqrt{m\omega}}.
\label{9}
\end{eqnarray}

Let us discuss the asymptotic behaviour of the possible solutions to Eq. (\ref{8}), which are determined for $r\rightarrow0$ and $r\rightarrow\infty$. From Refs. \cite{eug,mhv,vercin}, the behaviour of the possible solutions to Eq. (\ref{8}) at $r\rightarrow0$ and $r\rightarrow\infty$ allows us to write the function $R\left(r\right)$ in terms of an unknown function $H\left(r\right)$ as follows:
\begin{eqnarray}
R\left(r\right)=e^{-\frac{r^{2}}{2}}\,r^{\left|\gamma\right|}\,H\left(r\right).
\label{10}
\end{eqnarray}
Substituting Eq. (\ref{10}) into Eq. (\ref{8}), we obtain
\begin{eqnarray}
\frac{d^{2}H}{dr^{2}}+\left[\left(2\left|\gamma\right|+1\right)\frac{1}{r}-2r\right]\frac{dH}{dr}+\left[\frac{\beta}{m\omega}-2-2\left|\gamma\right|+\frac{\delta}{r}\right]H=0.
\label{11}
\end{eqnarray}

The second order differential equation (\ref{11}) corresponds to the biconfluent Heun equation \cite{heun,eug,b50} and the function $H\left(r\right)$ is the biconfluent Heun function: $H\left(r\right)=H_{\mathrm{B}}\left(2\left|\gamma\right|,\,0,\,\frac{\beta}{m\omega},\,2\delta,-r\right)$.
%\begin{eqnarray}
%H\left(r\right)=H\left(2\left|\gamma\right|,\,0,\,\frac{\beta}{m\omega},\,2\delta,-r\right).
%\label{12}
%\end{eqnarray}  
In order to proceed with our discussion about bound states solutions, let us use the Frobenius method \cite{arf,f1}. Thereby, the solution to Eq. (\ref{11}) can be written as a power series expansion around the origin:
\begin{eqnarray}
H\left(r\right)=\sum_{j=0}^{\infty}\,a_{j}\,r^{j}.
\label{13}
\end{eqnarray} 

Substituting the series (\ref{13}) into (\ref{11}), we obtain the recurrence relation:
\begin{eqnarray}
a_{j+2}=-\frac{\delta}{\left(j+2\right)\,\left(j+1+\alpha\right)}\,a_{j+1}-\frac{\left(g-2j\right)}{\left(j+2\right)\,\left(j+1+\alpha\right)}\,a_{j},
\label{14}
\end{eqnarray}
where
\begin{eqnarray}
\alpha=2\left|\gamma\right|+1;\,\,\,\,g=\frac{\beta}{m\omega}-2-2\left|\gamma\right|.
\label{14a}
\end{eqnarray}

By starting with $a_{0}=1$ and using the relation (\ref{14}), we can calculate other coefficients of the power series expansion (\ref{13}). For instance,
\begin{eqnarray}
a_{1}&=&-\frac{\delta}{\alpha}\nonumber\\
&=&-\frac{2\mathcal{E}f}{\sqrt{m\omega}}\frac{1}{\left(2\left|\gamma\right|+1\right)};\nonumber\\
[-2mm]\label{15}\\[-2mm]
a_{2}&=&\frac{\delta^{2}}{2\alpha\left(1+\alpha\right)}-\frac{\theta}{2\left(1+\alpha\right)}\nonumber\\
&=&\frac{2\mathcal{E}^{2}f^{2}}{m\omega}\frac{1}{\left(2\left|\gamma\right|+1\right)\left(2\left|\gamma\right|+2\right)}-\frac{g}{2\left(2\left|\gamma\right|+2\right)}.\nonumber
\end{eqnarray}

The quantum theory requires that the wave function (\ref{5}) must be normalizable, then, we assume that the function $R\left(r\right)$ vanishes at $r\rightarrow0$ and $r\rightarrow\infty$. In this way, bound state solutions can be obtained because there is no divergence of the wave function at $r\rightarrow0$ and $r\rightarrow\infty$. On the other hand, we have written the function $H\left(r\right)$ as a power series expansion around the origin in Eq. (\ref{13}). Thereby, bound state solutions can be achieved by imposing that the power series expansion (\ref{13}) or the biconfluent Heun series becomes a polynomial of degree $n$. In this way, we guarantee that $R\left(r\right)$ behaves as $r^{\left|\gamma\right|}$ at the origin and vanishes at $r\rightarrow\infty$ \cite{vercin,mhv}. Through the recurrence relation (\ref{14}), we can see that the power series expansion (\ref{13}) becomes a polynomial of degree $n$ by imposing two conditions \cite{f1,bb2,bb4,eug,b50,vercin,mhv,bf}:
\begin{eqnarray}
g=2n\,\,\,\,\,\,\mathrm{and}\,\,\,\,\,\,a_{n+1}=0,
\label{16}
\end{eqnarray}
where $n=1,2,3,\ldots$ and $g$ is given in Eq. (\ref{14a}). From the condition $g=2n$, we can obtain the expression for the energy levels for bound states:
\begin{eqnarray}
\mathcal{E}_{n,\,l}^{2}=m^{2}+m\,\omega_{n,\,l}\left[2n+2\left|\gamma\right|+1\right].
\label{17}
\end{eqnarray}

Hence, by introducing the scalar potential via minimal coupling, then, we can see in Eq. (\ref{17}) that the relativistic energy levels of the Klein-Gordon oscillator is modified by the influence of the Coulomb potential. This influence yields the ground state of the Klein-Gordon oscillator to be defined by the quantum number $n=1$ in contrast to the quantum number $n=0$ as obtained in Ref. \cite{kgo}. 

%Besides, the angular frequency of the Klein-Gordon oscillator is written in terms of the quantum numbers $\left\{n,\,l\right\}$ due to the exact solutions to Eq. (\ref{11}) can only be obtained for some values of the angular frequency. This restriction of the values of the angular frequency of the Klein-Gordon oscillator arises from the influence of the Coulomb potential on this relativistic oscillator. 

%Let us discuss this behaviour of the angular frequency from now on. This analysis comes from 

Let us now analyse the condition $a_{n+1}=0$ imposed in Eq. (\ref{16}) in order to obtain a polynomial of degree $n$ the power series expansion given in Eq. (\ref{13}). Thereby, let us assume that the angular frequency $\omega$ of the Klein-Gordon oscillator can be adjusted in such a way that the condition $a_{n+1}=0$ is satisfied. As a consequence, the quantum numbers of the system  $\left\{n,\,l\right\}$ restrict the possible values of the angular frequency. Therefore, there are values of the angular frequency which are not allowed in the system. For this reason, we have labelled $\omega=\omega_{n,\,l}$ in Eq. (\ref{17}). In this way, the conditions established in Eq. (\ref{16}) are satisfied and a polynomial solution to the function $H\left(r\right)$ given in Eq. (\ref{13}) is achieved \cite{eug}. As an example, let us consider the ground state $n=1$ and analyse the condition $a_{n+1}=0$. For $n=1$, thus, the condition $a_{n+1}=0$ yields $a_{2}=0$. By using Eq. (\ref{15}), then, the condition $a_{2}=0$ yields 
\begin{eqnarray}
\omega_{1,\,\,l}=\frac{2\,\mathcal{E}_{1,\,l}^{2}\,\,f^{2}}{m}\frac{1}{\left(2\left|\gamma\right|+1\right)},
\label{18}
\end{eqnarray}
which corresponds to the possible values of the angular frequency of the Klein-Gordon oscillator in the ground state. By substituting Eq. (\ref{18}) into Eq. (\ref{17}), the energy levels of the ground state is given by
\begin{eqnarray}
\mathcal{E}_{1,\,l}=\pm \frac{m}{\sqrt{1-2f^{2}\frac{\left(3+2\left|\gamma\right|\right)}{\left(2\left|\gamma\right|+1\right)}}}.
\label{19}
\end{eqnarray}

 In what follows, let us consider the simplest case of the function $H\left(r\right)$ which corresponds to a polynomial of first degree. In this way, for $n=1$, we can write $H_{1,\,l}\left(r\right)=1+\frac{\delta}{\alpha}\,r$. Thereby, the radial wave function (\ref{10}) associated with the ground state is given in the form: 
\begin{eqnarray}
R_{1,\,l}\left(r\right)=e^{-r^{2}/2}\,r^{\left|\gamma\right|}\,\left(1+\frac{\delta}{\alpha}\,r\right).
\label{19}
\end{eqnarray}

Therefore, from the introduction of the scalar potential via minimal coupling in the Klein-Gordon equation, we have that the effects of the Coulomb potential on the the spectrum of energy of the Klein-Gordon oscillator is given by a change of the energy levels, where the ground state is defined by the quantum number $n=1$. Moreover, the values of the angular frequency of the Klein-Gordon oscillator are restricted to a set of values in which allow us to obtain a polynomial solution to the biconfluent Heun series. From the the quantum mechanics point of view, this is an effect characterized by the dependence of angular frequency of the Klein-Gordon oscillator on the quantum numbers $\left\{n,l\right\}$ of the system \cite{eug,b50}.

\section{Klein-Gordon oscillator under the influence of a Coulomb potential and a linear scalar potential}

Let us extend our discussion by introducing a scalar potential into the Klein-Gordon equation by modifying in the mass term in the form: $m\rightarrow m+V\left(\vec{r},\,t\right)$, where $V\left(\vec{r},\,t\right)$ is the scalar potential \cite{scalar,greiner}. Let us consider a linear scalar potential given by:
\begin{eqnarray}
V\left(\rho\right)=\nu\,\rho,
\label{3.1}
\end{eqnarray}
where $\nu$ is a constant that characterizes the linear confining potential. It has been proposed to describe the confinement of quarks \cite{linear,linear1} due to experimental data show a behaviour of the confinement to be proportional to the distance between the quarks \cite{linear4,linear4a,linear4b,linear4c}. It has also been explored in studies of the quark-antiquark interaction as a problem of a relativistic spinless particle which possesses a position-dependent mass, where the mass term acquires a contribution given by a interaction potential that consists of a linear and a harmonic confining potential plus a Coulomb potential term \cite{bah}. Furthermore, the linear scalar potential has attracted a great interest in atomic and molecular physics as pointed out in Refs. \cite{linear3a,linear3b,linear3c,linear3d,linear3e,linear3f} and in several discussions of relativistic quantum mechanics \cite{linear2,linear2a,linear2b,linear2c,linear2d,linear2e,linear2f,eug,scalar2,vercin,mhv}.

Hence, the general form of the Klein-Gordon equation describing the interaction of the Klein-Gordon oscillator with the static scalar potential (\ref{3.1}) and the Coulomb potential (\ref{1.3}) is given by (with $c=\hbar=1$)
\begin{eqnarray}
\left[m+V\right]^{2}\phi=\left[i\frac{\partial}{\partial t}+q\,A_{0}\right]^{2}\phi-\left[\hat{p}+im\omega\rho\,\hat{\rho}\right]\cdot\left[\hat{p}-im\omega\rho\,\hat{\rho}\right]\phi.
\label{3.2}
\end{eqnarray}
By substituting Eqs. (\ref{3.1}) and (\ref{1.3}), then, Eq. (\ref{3.2}) becomes
\begin{eqnarray}
m^{2}\phi&=&-\frac{\partial^{2}\phi}{\partial t^{2}}+i\frac{2f}{\rho}\frac{\partial\phi}{\partial t}+\frac{f^{2}}{\rho^{2}}\,\phi+\frac{\partial^{2}\phi}{\partial\rho^{2}}+\frac{1}{\rho}\frac{\partial\phi}{\partial\rho}+\frac{1}{\rho^{2}}\frac{\partial^{2}\phi}{\partial\varphi^{2}}\nonumber\\
[-2mm]\label{3.3}\\[-2mm]
&+&m\omega\,\phi-m^{2}\omega^{2}\rho^{2}\,\phi-2m\nu\,\rho\,\phi-\nu^{2}\,\rho^{2}\,\phi.\nonumber
\end{eqnarray}

By following the steps from Eq. (\ref{1.5}) to Eq. (\ref{7}), we obtain the radial equation
\begin{eqnarray}
\frac{d^{2}R}{d\rho^{2}}+\frac{1}{\rho}\,\frac{dR}{d\rho}-\frac{\gamma^{2}}{\rho^{2}}\,R+\frac{2f\mathcal{E}}{\rho}\,R-2m\nu\,\rho\,R-\theta^{2}R+\beta\,R=0,
\label{3.4}
\end{eqnarray}
where the parameters $\beta$ and $\gamma$ are defined in Eq. (\ref{7}) and the parameter $\theta$ is defined as
\begin{eqnarray}
\theta^{2}=m^{2}\omega^{2}+\nu^{2}.
\label{3.5}
\end{eqnarray}

Let us perform a change of variables given by $\xi=\sqrt{\theta}\,\rho$, then, the radial equation (\ref{3.4}) becomes
\begin{eqnarray}
\frac{d^{2}R}{d\xi^{2}}+\frac{1}{\xi}\,\frac{dR}{d\xi}-\frac{\gamma^{2}}{\rho^{2}}\,R+\frac{\tau}{\xi}\,R-\mu\,\xi\,R-\xi^{2}\,R+\frac{\beta}{\theta}\,R=0,
\label{3.6}
\end{eqnarray}
where we have defined the parameters
\begin{eqnarray}
\tau=\frac{2f\mathcal{E}}{\sqrt{\theta}};\,\,\,\,\,\mu=\frac{2m\nu}{\theta^{3/2}}.
\label{3.7}
\end{eqnarray}

By analysing the asymptotic behaviour of the possible solutions to Eq. (\ref{3.6}) as in the previous section, we have that we can write the function $R\left(\xi\right)$ in the form:
\begin{eqnarray}
R\left(\xi\right)=e^{-\xi^{2}/2}\,e^{-\mu\xi/2}\,\xi^{\left|\gamma\right|}\,\bar{H}\left(\xi\right),
\label{3.8}
\end{eqnarray}
where $\bar{H}\left(\xi\right)$ is an unknown function. After substituting Eq. (\ref{3.8}) into Eq. (\ref{3.6}), we obtain the following equation for the function $\bar{H}\left(\xi\right)$:
\begin{eqnarray}
\frac{d^{2}\bar{H}}{d\xi^{2}}+\left[\frac{\left(2\left|\gamma\right|+1\right)}{\xi}-\mu-2\xi\right]\frac{d\bar{H}}{d\xi}+\left[\sigma+\frac{\vartheta}{\xi}\right]\bar{H}=0.
\label{3.9}
\end{eqnarray}
where we have established that 
\begin{eqnarray}
\sigma=\frac{\beta}{\theta}+\frac{\mu^{2}}{4}-2-2\left|\gamma\right|;\,\,\,\,\vartheta=\frac{\mu}{2}\left(2\left|\gamma\right|+1\right)+\tau.
\label{3.10}
\end{eqnarray}

Observer that Eq. (\ref{3.9}) is also a biconfluent Heun equation \cite{heun} and the function $\bar{H}\left(\xi\right)=H_{\mathrm{B}}\left(2\left|\gamma\right|,\,\mu,\,\frac{\beta}{\theta}+\frac{\mu^{2}}{4},\,-2\tau,\,\xi\right)$ is the biconfluent Heun function. By following the steps from Eq. (\ref{13}) to Eq. (\ref{16}), we obtain the following recurrence relation:
\begin{eqnarray}
a_{j+2}=\frac{\mu\left(j+1\right)+\vartheta}{\left(k+2\right)\left(k+2+2\left|\gamma\right|\right)}\,a_{j+2}-\frac{\sigma-2j}{\left(k+2\right)\left(k+2+2\left|\gamma\right|\right)}\,a_{j}.
\label{3.11}
\end{eqnarray}
By starting from $a_{0}=1$ as in the previous section, the coefficients $a_{1}$ and $a_{2}$ becomes
\begin{eqnarray}
a_{1}&=&\frac{\mu}{2}+\frac{\tau}{\left(1+2\left|\gamma\right|\right)};\nonumber\\
[-2mm]\label{3.12}\\[-2mm]
a_{2}&=&\frac{\vartheta\left(\vartheta+\mu\right)}{2\left(1+2\left|\gamma\right|\right)\left(2+2\left|\gamma\right|\right)}-\frac{\sigma}{2\left(2+2\left|\gamma\right|\right)}.\nonumber
\end{eqnarray}

Now, we have that the series (\ref{13}) becomes a polynomial of degree $n$ by imposing that:
\begin{eqnarray}
\sigma=2n;\,\,\,\,a_{n+1}=0;
\label{3.13}
\end{eqnarray}
where $n=1,2,3,\ldots$. From the condition $\sigma=2n$, we obtain
\begin{eqnarray}
\mathcal{E}_{n,\,l}^{2}=m^{2}-m\omega+2\theta\left[n+\left|\gamma\right|+1\right]-\frac{m^{2}\nu^{2}}{\theta^{2}}.
\label{3.14}
\end{eqnarray}

On the other hand, by analysing the condition $a_{n+1}=0$ for the ground state ($n=1$) as in the previous section, we obtain a third degree algebraic equation given by
\begin{eqnarray}
\theta_{1,\,l}^{3}-\frac{2f^{2}\mathcal{E}_{1,\,l}^{2}}{\left(2+2\left|\gamma\right|\right)}\,\theta_{1,\,l}^{2}-2m\nu\,f\,\mathcal{E}_{1,\,l}\,\theta_{1,\,l}-m^{2}\nu^{2}\,\frac{\left(1+2\left|\gamma\right|\right)\left(3+2\left|\gamma\right|\right)}{2\left(2+\left|\gamma\right|\right)}=0;
\label{3.15}
\end{eqnarray}
where we have labelled $\theta_{n,\,l}=\sqrt{m^{2}\,\omega_{n,\,l}^{2}+\nu^{2}}$ in order to establish that we are considering the angular frequency of the Klein-Gordon oscillator to be the parameter that can be adjusted in order that the condition $a_{n+1}=0$ can be satisfied and a polynomial solution to $\bar{H}\left(\xi\right)$ can be achieved. Since Eq. (\ref{3.15}) has at least one real solution \cite{eug}, then, the expression for the energy level of the ground state $\mathcal{E}_{1,\,l}$ can be obtained from this real solution. However, we do not write it here because the real solution of Eq. (\ref{3.15}) is very long. Observe that, for other energy levels, different equations for $\theta_{n,\,l}$ or $\omega_{n,\,l}$ can be obtained from the condition $a_{n+1}=0$.

\section{Conclusions}

We have seen that the effects of the Coulomb potential on the Klein-Gordon oscillator is given by the modification of the spectrum of energy of the Klein-Gordon oscillator, where the ground state is defined by the quantum number $n=1$ instead of the quantum number $n=0$ as given in Ref. \cite{kgo} and the values of the angular frequency are restricted and depend on the quantum numbers $\left\{n,\,l\right\}$ of the system. The meaning of this restriction of the values of the angular frequency is that only a set of values allow us to obtain a polynomial solution to the biconfluent Heun series. As an example, we have calculated the angular frequency of the ground state $n=1$ and obtained the expression of the energy level of the ground state.

We have also analysed the influence of a linear scalar potential and the Coulomb potential on the Klein-Gordon oscillator. We have seen that the energy levels of the Klein-Gordon oscillator are modified, where the ground state is also determined by the quantum number $n=1$ instead of the quantum number $n=0$, and possible values of the angular frequency are determined by the quantum numbers $\left\{n,\,l\right\}$ of the system. In particular, we have shown that the possible values of the angular frequency of the Klein-Gordon oscillator associated with the ground state of the system are determined by a third degree algebraic equation.

Recently \cite{bf}, we have investigated the the effects of the Coulomb-type potential introduced by a coupling with the mass term on the spectrum of energy of the Klein-Gordon oscillator. Despite the Klein-Gordon oscillator is under the influence of a scalar potential, these two couplings yield different results, for instance, for the angular frequency $\omega_{1,\,l}$ and the relativistic energy level $\mathcal{E}_{1,\,l}$ associated with the ground state of the Klein-Gordon oscillator. It is worth mentioning that the influence of the Klein-Gordon oscillator and the Coulomb potential on a scalar particle can be of interest in studies of the quark-antiquark interaction \cite{bah}, the Kaluza-Klein theory \cite{furtado}, the Casimir effect \cite{casimir,mb1} and relativistic effects on condensed matters systems such as effects associated with linear topological defects in solids \cite{kleinert,kat,moraesG2}, the Aharonov-Bohm effect for bound states \cite{pesk,fur5,ab7} and persistent currents \cite{by,tan}.

\acknowledgments{The authors would like to thank the Brazilian agencies CNPq and CAPES for financial support.}

\end{document}